\documentstyle[parkcity,pslatex,rawfonts]{article}  
\frompage{000} \topage{000}                                              
\def\rightmark{Stopping in HI reactions} 
\def\leftmark{F.~Videb\ae k}

\newcommand{\nc}{\newcommand}
\nc{\bce}{\begin{center}}
\nc{\ece}{\end{center}}
\nc{\dyav}{$\delta y_{p}$}
\nc{\snn}{$\sqrt{s_{NN}}~$}
\nc{\ppbar}{N$(\bar{\rm p}$)/N(p)~}
\nc{\gevc}{GeV/$c$}
\nc{\pbar}{$\bar{\rm p}$}

\begin{document}
 
\title{Stopping in Relativistic Heavy Ion Reactions \\From SIS to RHIC} 
\authors{
{Flemming Videb\ae k$^1$
}\\[2.812mm]
{\normalsize
\hspace*{-8pt}$^1$Physics Department,\\
Brookhaven National Laboratory, \\ 
NY 11973, USA\\[0.2ex] 
}}
 
\abstract{The status of stopping in heavy ion reactions is reviewed
by comparing available data in pp, pA and AA systems over the energy regime
$\sqrt{s_{NN}} \approx 2.5 - 130$ GeV. The data consist of average rapidity losses, anti-proton over proton ratios,
and net-baryon values at mid-rapidity. The overall features of nuclear stopping are reasonbly well described by simple
extrapolations of pp, and pA collisions to AA.}

\keyword{Nuclear Stopping, Baryon Distributions.}
\PACS{25.75.-q}

\maketitle

\section{Introduction}\label{intro}

The major goal for relativistic heavy ion reactions is to form hot and dense nuclear matter, 
to study its properties, and to come to a better understanding of 
non-pertubative QCD. 
The heavy ion experiments at RHIC have this as their main purpose.
It  is important to understand nuclear stopping in detail since it is a requisite for
the formation of such systems in describing the conversion of the initial kinetic 
energy into matter excitation at mid-rapidity. A description in detail of how this transport
takes place is a necessary ingredient in our overall understanding of the reactions, and in the
expectations to form and study the properties of QGP.

The importance of stopping was recognized early \cite{Bus84,Date85} with a good conceptual understanding
of pA and AA stopping. This also laid the foundation for estimations of what energy 
densities might be reached in heavy ion induced reactions. Experimental data have come 
at a slow but steady pace as the accelerators and experiments have been build and data analyzed.
Some years back I surveyed  the available data with Ole Hansen \cite{Vid95}, and now is a good time to revisit 
the subject, since more detailed data has emerged at lower energies, 
not only in AA collissions but also in pp and pA. 
In addition, the first run at RHIC in 2000 has given us important  clues about
the de facto situation on stopping at extreme ultra relativistic energies. In this paper I will
review some recent experimental data on stopping coming from pp, pA at lower  energies, 
as well as data extending our knowledge about AA.
I will put the data into perspective by comparing rapidity losses with
estimates from the multichain model \cite{Date85} for the energy dependence. 
Though the first data from RHIC did not measure baryon rapidity distributions 
and thus enable us to quantify the amount of stopping, the particle production at mid-rapidity 
and the anti-proton to proton ratios are indicator of stopping  and is discussed in the last section.

\section{pp and pA reactions}
It has often been conjectured that a thorough understanding of pp and pA reactions is needed
to gauge the significance of results from AA, whether results can be interpreted as merely
simple superpositions of more elementary collisions or if other many-body effects play a role.
Such superpositions or extrapolations clearly rely on high quality data from pp and pA. 
The data that formed the basis for the early understanding were in many ways limited, 
in large part due to lack of statistics and lack of centrality event selection. 
I will not review these here, but merely point to the references 
given in the earlier review\cite{Bus84,Vid95,Bus88,Her99}.

\begin{figure}[h]
\vspace*{-0.2cm}
\bce
\epsfysize 6.cm      \epsfbox{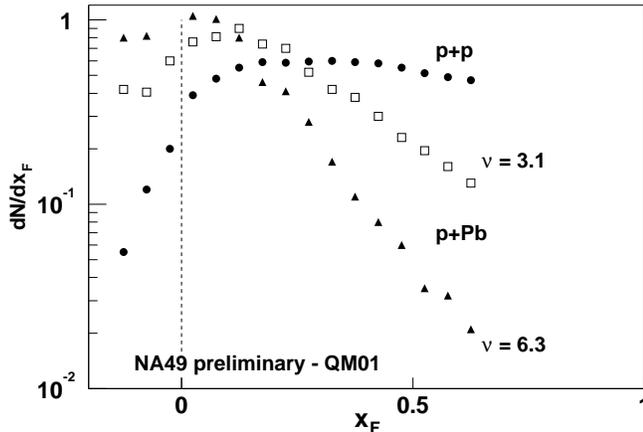}
\vspace*{-0.5cm}
  \caption{
    dN/dx$_{f}$ distribution for pp, and pPb from  preliminary  NA49 data \cite{Blu01}.}
  \label{fig:dndxf}
\ece
\end{figure}   

In recent years significant data have emerged from experiments at AGS and CERN. In particular the data
from E910, E941 at the AGS, and NA49 at the SPS are most significant in terms of rapidity coverage and centrality selection. 
The data have yet to be published, but preliminary results have been presented at the Quark Matter 
2001 conference in Stony Brook
\cite{Blu01,Cole01,Tai01}, and at this conference \cite{Seyboth01}.
In general the data have been selected on centrality by counting 'grey' tracks i.e. slow protons emitted from the target. 
The 'grey' tracks have been shown by model calculations to be related to the number of collisions 
that the incoming proton suffers in the nucleus. 
The data show that  after the first collision the system has forgotten the isospin content of the entrance channel 
as manifested by the observations that spectra, 
and rapidity distributions of pions, protons and neutrons are similar \cite{Cole01} 
when gated on number of collisions greater than one.

The NA49 experiment has made further progress by isolating the behavior of the stopped projectile by nucleus. 
This is done by subtracting the component from the target measuring the protons in $\pi$--A reaction. 
The pion has no baryon content, and the produced
protons  correspond to the target contribution in pA reactions. 
The preliminary dN/d$x_{\rm F}$ distribution are shown in Fig.~\ref{fig:dndxf}. The 
Feynman $x_{\rm F}$ is defined in the interval -1 to 1 as P$_L$(c.m.)/P$_L$(c.m.)$_{max}$. 
The dN/dx$_{\rm F}$ for pp is almost constant, as has been known for a long time.
The pPb data are shown for two values of the mean number $\nu$ of collisions, 
$\nu = 3.1$ corrsponding to mean bias pA and $\nu = 6.3$
to central collisions. 
The central collision events exhibit complete stopping with the peak of the distribution at x$_{\rm F}$ =0. 
These detailed
measurements confirm the previous expectations of Busza and Goldhaber \cite{Bus84}.
NA49 also measured distributions of neutrons emitted in pp and pA and found the distributions after one
collision to be nearly identical to those of stopped protons.
\begin{figure}[htp]	
\bce
\epsfysize 6.cm      \epsfbox{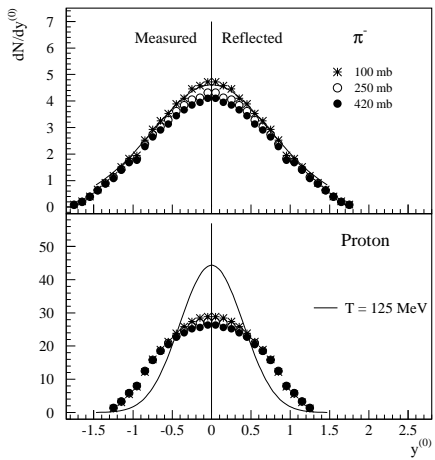}	
  \caption{
   Pion and proton  dN/dy distributions in NiNi reactions at 1.93AGeV from the FOPI Collaboration \cite{Hon98}.}
  \label{fig:fopi}
\ece
\end{figure}    

\section{AA Collisions}

An example of the data obtained since the previous review is shown in Fig.~\ref{fig:fopi}, the rapidity distributions 
of pions and protons in NiNi reactions at 1.93A GeV \cite{Hon98}. The data are for three very central collision bins. 
The most central 100 mb cross section bin show a slight enhanchement both of pion production and of protons, indicative of a slight
increase in stopping. 
The data are compared to the expectation of a pure isotropic Boltzman distribution with a temparature of T=125 MeV.
Already at this low energy we do not obeserve complete stopping in the sense of isotropic decay of a fireball, but see effects
of incomplete stopping or of flow in terms of longitudinal expansion.

\begin{figure}[htp]
\vspace*{-0.5cm}
\bce
\epsfysize 6cm      

\epsfbox{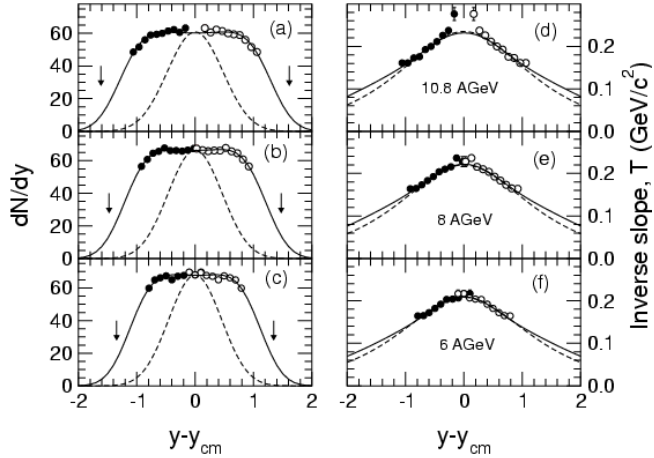}
  \caption{
    The left panel shows proton rapidity distribution at AGS energies from E917 \cite{E917proton}, and the right panels
the extracted temperatures from Boltzman fits to the data. }
  \label{fig:917}
\ece
\end{figure}

The E917 experiment has extended the earlier measurement of E866 \cite{E866proton} at 11.7 AGeV/$c$ with measurements at
6, 8 AGeV \cite{E917proton}. The proton rapidity distributions are displayed in the left panel of Fig.~\ref{fig:917}. Again the data 
are compared to expectation of isotropic decay, and incomplete stopping is observed. The data were analyzed in a novel way,
attributing  the overall distributions as being a sum of two Gaussians representing an incomplete stopping 
of projectile and target protons.  
Such decomposition attempts  to determine the projectile stopping as expressed by the average rapidity loss. The values
extracted by this method are similar to those obtained and described in the next paragraph.

For symmetric systems the average rapidity loss of protons/baryons are defined as
\begin{equation}
\delta y_{p} =  {\int dy dN/dy (y_{p}-y) \over \int dy dN/dy}
\end{equation}
The integral is taken from mid-rapidity (y= 0) to the beam rapidity $y_{p}$ and is an approximation to the 
actual projectile projectile, since it is not possible 
to distinguish between target and projectile protons when protons pile up at mid-rapidity.
\begin{figure}[htp]
\vspace*{-0.5cm}
\bce
\epsfysize 6cm      \epsfbox{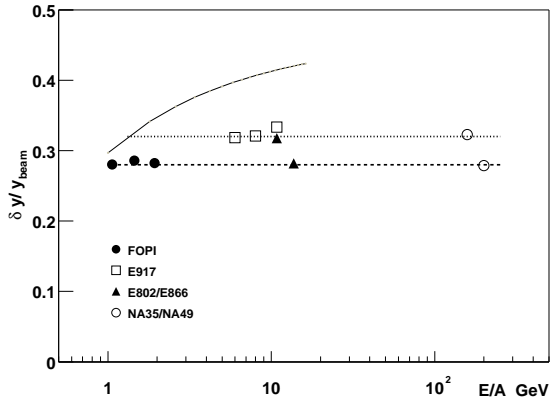}
\vspace*{-0.7cm}
  \caption{
Average rapidity relative to the beam rapidity vs. laboratory energy. 
The experimental data are for light systems SS SiAl, and NiNi, and heavy AuAu and PbPb,
from the FOPI, E802, E866, E917 and NA49 experiment. The NA49 data are from Ref.\cite{Appl99}. The reference for the other data are given in the text.
}
  \label{fig:dyexp}
\ece
\end{figure}    
Figure \ref{fig:dyexp} symmetrizes the results of AA average rapidity loss \dyav ~relative to the beam rapidity versus
the kinetic energy per nucleon. With the additional data available it is apparent that both the heavier systems and the lighter systems
both have constant relative rapidity losses  with the heavier slightly higher with an average value of 0.32. Such higher value is of course
not unexpected due to the increase in thickness of the nuclei involved.
Such plot is also deceiving in displaying no energy dependence.  The absolute value does increase with energy. and so the amount
of available energy for particle production.
The expectation of \dyav ~from a isotropic Boltzman distribution with a temperature of 125 MeV is shown on the figure rising from
about 0.30 at the lowest energy, again illustrating that such isotropic stopping is not observed at relativistic energies.

\section{Multichain Model}
The multichain model was introduced in Ref.\cite{Date85} to study stopping in pA and AA reactions.
This model is used to illustrate how well the overall features of nuclear stopping is described by such simple
approaches. Similar analysis could also be carried out using one of the  many event generator available. 
These models have quite similar assumptions.
The basic assumption and points are
\begin{itemize}
\item{} The nuclear geometry is seperated from the dynamics.
\item{} Transverse distribution are independent of E,A, and p$_{L}$.
\item{} Glauber or wounded nucleon model is used to calculate collision probabilities.
\item{} Multiple collision dynamic scales with Feynman x.
\item{} Independent projectile and target fragmentation.
\end{itemize}

The calculations presented here utilize equations presented in section 2 and 5 of Ref.~{\cite{Date85}} with a constant
fragmentation function for pp, and and a value of $\alpha$=1.3 at the lower energies and 2.7 at the higher. Such value was derived in
\cite{Tai01}  for AGS and SPS energies, respectively. Since the calculation here is done down to low values of \snn,  both target
and projectile contribution were included,  and at cutoff was introduced in the allowed final $x_{\rm F}$ 
to require open inelastic channels.

\begin{figure}[htp]
\vspace*{-0.5cm}
\bce
\epsfysize 6.cm      \epsfbox{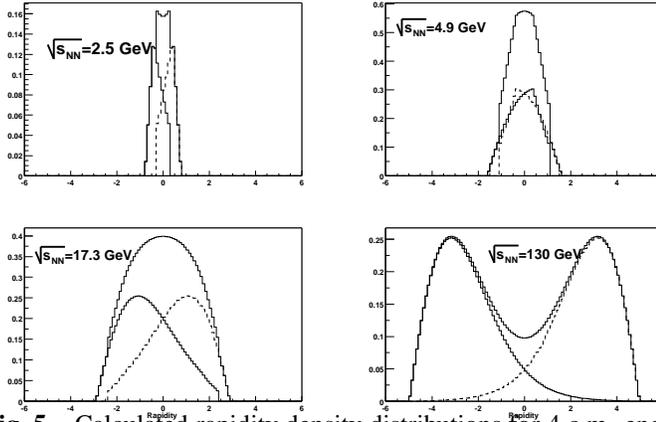}
\vspace*{-0.7cm}
  \caption{
Calculated rapidity density distributions for 4 c.m. energies using the multichain model.}
  \label{fig:dycalc}
\ece
\end{figure}    

\begin{figure}[htp]
\vspace*{-0.5cm}
\bce
\epsfysize 6.cm      \epsfbox{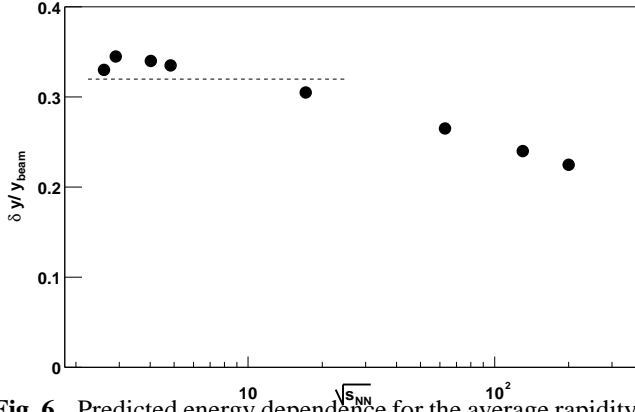}
\vspace*{-0.7cm}
  \caption{
    Predicted energy dependence for the average rapidity loss
in the multichain model. 
The dashed line shows the near constant value observed in AA collisions at lower energies.}
  \label{fig:dyvss}
\ece
\end{figure}

This relatively simple model allows us to make predictions of the rapidity loss vs c.m. energy.
The results from this calculation, assuming a standard Wood-Saxon distribtuion of the nuclear densities, 
and an inelastic nucleon-nucleon cross section of 30mb 
at the lower energies and 40 mb at RHIC energies are displayed in Fig.~\ref{fig:dycalc}. 
The 4 panels show un-normalized rapidity distributions for the projectile 
and target baryons as the dashed lines, while the total distributions are given as solid 
lines at four c.m. energies from SIS to RHIC. The evolution from overlap between projectile and target,
 and nearly full stopping to partial transperency is clearly observed. The average rapidity loss \dyav 
~is calculated from the distributions and the plotted relative to the beam rapidity in Fig.~\ref{fig:dyvss} versus the c.m. energy. 
This schematic model does in fact quite accurately describe the observed near
constant value in the lower energy data shown in Fig.\ref{fig:dyexp}.
Similar features are also predicted from e.g. cascade code, but the present model analysis demonstrates
that the main features are described starting with the elementary pp energy loss mechanism, 
and folding with a geometric description of heavy ion reactions using a Glauber or Wounded Nucleon Model.

\section{Early Results from RHIC}

Most of the early results from RHIC have been focused at mid-rapidity, determining charged particle densities,
and for identified particles looking at ratios to minimize the systematic errors. Thus measurement of
net-baryon number densities at mid-rapidity (though preliminary numbers have been quoted by STAR), 
nor a determination of the rapidity distributions of net-baryons from which rapidity losses can be determined has been made.
Even so these measurements shed light on stopping at \snn=130 ~GeV ~by investigating 
the ratio \ppbar at mid-rapidity and its centrality and $p_t$ dependence. So far 3 experiments have reported final values of \ppbar (see Refs~
\cite{STAR01,BRAHMSppbar,PHOBOSppbar}).
The values all agree and are in the range of $0.57-0.65$ for central collisions near $y=0$. It is of interest to compare this value
with heavy ion data from lower energies. Data exist for Au-Au collisions at the AGS \cite{Ahle98},  near threshold for \pbar ~production,
and at SPS~\cite{SPSppbar}. Data at higher energies have been measured in pp reactions at the ISR~\cite{Guettler76,Capi74}. 
These later 
data cannot directly compared with the heavy ion data, but require an iso-spin correction. A simple estimate for such correction is
described in the appendix. The available data are displayed in Fig.\ref{fig:ppbar}. The ISR data are displayed both with the measured values
(open symbols) 
as well as with the isospin corrected (filled symbols). An overall smooth dependence with \snn~ is observed with the 
RHIC heavy ion data being the largest. 
The values observed at \snn=130GeV~ are also close to an extrapolation of the \ppbar ratios from ISR~\cite{Boggild01}.

\begin{figure}[htp]
\vspace*{-0.5cm}
\bce
\epsfysize 6.cm      \epsfbox{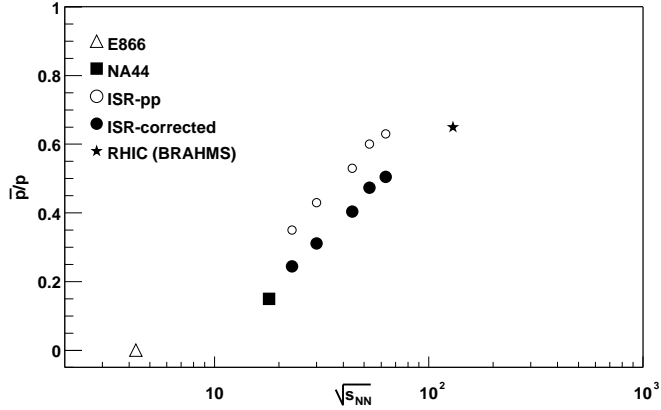}
\vspace*{-0.2cm}
  \caption{
    Energy dependence of $\bar{{\rm p}}/{\rm }p$ ratio from AA and pp data.
}
  \label{fig:ppbar}
\ece
\end{figure}    

\begin{figure}[htp]
\vspace*{-0.5cm}
\bce
\epsfysize 6.cm      \epsfbox{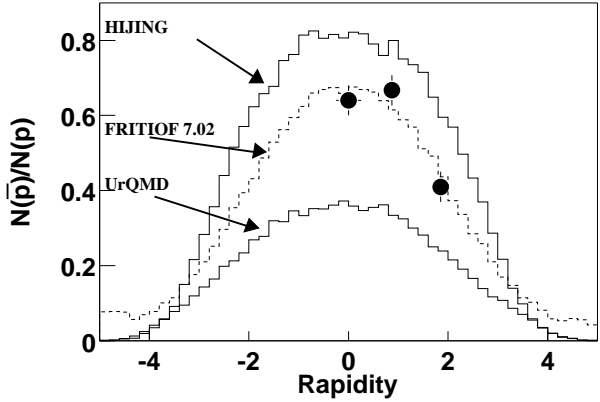}
\vspace*{-1.0cm}
  \caption{
    Comparison of the measured  \ppbar  ratios to model
    predictions (BRAHMS data). The data shown are for $0-40\%$ central events.
    The three model calculations (HIJING, FRITIOF, and UrQMD) are shown for comparison. See text for details.} 
  \label{fig:brahmspp}
\ece
\end{figure}

BRAHMS has measured \ppbar not only at y=0, but also at two larger rapidity values \cite{BRAHMSppbar}, 
which shed further light on the issue of stopping.
The rapidity dependence of the \ppbar  ratio is shown in Fig.~\ref{fig:brahmspp}. 
The present systematics  show that the
ratios fall off more rapidly over two units of rapidity  than the
nucleus-nucleus results at lower  $\sqrt{s_{NN}}$, but are very
similar to the p+p result at roughly half the CM energy
\cite{Capi74}. 

Also in Fig.~\ref{fig:brahmspp} is compared the measured ratios to
calculations using the HIJING model ~\cite{HIJING}, the FRITIOF
7.02~\cite{FRITIOF} string model  and the UrQMD  cascade model
~\cite{UrQMD} using the same cuts on centrality as those applied in the data analysis. 
The FRITIOF model reproduces the \ppbar  ratios quite well, while it overpredicts
by $\approx 30\%$ the charged particle yield at $\eta \approx 0$ ~\cite{Phobos00}.
This is mainly related to a fairly large degree of stopping predicted by the model.
On the other hand the HIJING model descibes the overall charge particle 
yields at $\eta\approx0$ while it fails in describing rapidity dependence of the anti-proton to proton 
ratio. This feature of the model is in turn related to a small
stopping of the projectile baryons.
The UrQMD model, which is not a partonic model, underpredicts the ratios by almost a factor of two.
These models exemplify the present understanding of heavy ion collisions at
this new energy regime; none of the models offers a consistent description of
the observed features, and the rapidity dependence serves as a benchmark for the description of
stopping and anti-baryon production.
 
\section{Conclusions}\label{concl}
The issue of stopping in heavy reactions has been studied in details 
in the past fifteen years and systematic heavy ion data on this exists from SIS to SPS energies, 
and will  emerge from the data at RHIC within the next year. 
The  rapidity losses have been measured  as functions of projectile, centrality, and energy. The existence of new 
high quality data from pp and pA will enable us to see if the AA stopping is merely 
simple extensions of pp and pA, or if new features due to the many-body overlaps manifest 
themselves in the details. 
The data reviewed here call for a careful theoretical analysis, and not a phenomenological approach.

\section{Appendix}\label{app}
An isospin correction is needed when comparing \ppbar ratios from pp with data from AA. The correction
arises  because the p multiplicity from pp$\rightarrow$ pX is 0.6, while for an isospin symmetric Nucleon-Nucleon (NN) 
system we expect NN$\rightarrow$ pX to have a multiplicity of 1.0. 
The first order correction is estimated as follows.
The proton cross section as coming from two components 
a) the direct stopping $\sigma_{S}$ and
b) a pair-production component $\sigma_{P}$ giving rise to an equal cross section of protons and \pbar's. 
The pp reaction measure $\sigma_{P}$ / ($\sigma_{P}$+$\sigma_{S}$) ,
while for the comparison with AA we want to estimate the same quantity with isospin weighted yields.   
This is achieved by replacing the $\sigma_{S}$  with a increased yield of +$\sigma_{S}/0.6$
In that case the isospin weighted ratios $\sigma_{\rm N}$ / ($\sigma_{P}$+$\sigma_{S}$)  is given by 
$\sigma_{P}$ / ($\sigma_{P}$+$\sigma_{S}/0.6$) with yields from the pp data.
The expected ratio for NN collisions can then be written as
\begin{equation}
N({\rm N})/N(\bar{\rm N}) = N({\rm p})/N(\bar{\rm p}) /(5/3-2/3*N({\rm p}/N(\bar{\rm p}))
\end{equation}
This has been used to correct the \ppbar ratios from ISR in Fig.\ref{fig:ppbar}.
\section*{Acknowledgements}

This work was supported by the Division of Nuclear Physics of the Office of Science of 
the U.S. Department of Energy under contract with BNL (No.DE-AC02-98CH10886). I value discussions
with Ole Hansen and H. B\o ggild on the  issue of stopping and data from ISR, and thank  F. Rami
for supplying me with the figure of the FOPI data, and J.H.Lee for comments to the manuscript.

\section*{Notes}
\begin{notes}
\item[a]
E-mail: videbaek@bnl.gov
\end{notes}

\vfill\eject

\end{document}